\documentclass[final,5p,times,twocolumn]{elsarticle}

\usepackage{epsfig}
\graphicspath{{Figures/}}
\usepackage{amssymb,amsmath}
\usepackage{xcolor}
\usepackage{lineno}

\journal{Journal of Magnetism and Magnetic Materials}  

\begin{document}
\begin{frontmatter}

\title{Correlation of uniaxial magnetic anisotropy axes and principal resistivities in polycrystalline ferromagnetic films}

\author[add1,add2]{Movaffaq Kateb}
\author[add2]{Snorri Ingvarsson\corref{cor1}}
\cortext[cor1]{Corresponding author: S. Ingvarsson (email: sthi@hi.is)}
\address[add1]{Department of Engineering, School of Technology, Reykjavik University, Menntavegi 1, IS-102 Reykjavik, Iceland}
\address[add2]{Science Institute, University of Iceland, Dunhaga 3, IS-107 Reykjavik, Iceland}

\begin{abstract}
In the present study, we demonstrate the measurement of resistivity tensor ($\boldsymbol{\rho}$) along the magnetic axes of a polycrystalline film of ferromagnetic permalloy (Py). To this end, conventional Hall-bar and a more recent extended van der Pauw methods were utilized for determining 2D $\boldsymbol{\rho}$ in the film plane. The samples were prepared by normal incidence sputter deposition within an \emph{in-situ} magnetic field to induce in-plane uniaxial magnetic anisotropy in the film. Since $\boldsymbol{\rho}$ might be affected by the internal magnetization of the film, we performed measurements by rotation of a saturating magnetic field in the film plane. Both methods indicate that the average resistivity is lower along the easy axis of the film compared to the hard axis. Since X-ray diffraction results indicated no dominating texture in the film, we concluded that there is a correlation between uniaxial magnetic axes and principal resistivity axes. This is an important finding that allows determining the direction of magnetic anisotropy axes without magnetometry. The results also verify atomic or pair ordering to be the origin of uniaxial magnetic anisotropy in the Py since resistivity is sensitive to the level of order in solids. The extended van der Pauw utilized here can be easily performed on the as-received samples which is of practical interest. 

\end{abstract}


\end{frontmatter}


\section{Introduction}
It has been shown that magnetic anisotropy is a tunable parameter through the alloy composition \citep{gupta2007}, atomic order \citep{kateb2019epi} or in ultra-thin layers through epitaxial strain and interface mixing \citep{kateb2019s,kateb2020}. Thus, various methods have been developed for tuning the magnetic anisotropy of ferromagnetic materials. However, the origin of induced uniaxial anisotropy, even in a popular ferromagnet such as permalloy Ni$_{80}$Fe$_{20}$ at.~\% (Py), has been a subject of huge debate over decades (cf.\ Ref.~\citep{kateb2019,kateb2019epi,kateb2018} and references therein). It has been thought that self-shadowing and off-normal texture are responsible for the uniaxial anisotropy induced by tilt deposition \citep{smith1959,smith1960}. We have already shown that uniaxial anisotropy can be achieved in very smooth (no-self-shadowing) Py films with normal texture obtained by tilt deposition \citep{kateb2019}.

A more conventional way of inducing uniaxial anisotropy is by post-annealing or growth in a magnetic field. Bozorth \cite{bozorth1934a,bozorth1934b} believed that applying an external magnetic field causes magnetostrictive deformation which during the annealing in a field becomes permanent. The uniaxial anisotropy induced by an \emph{in-situ} magnetic field during the growth can also be explained by the same interpretation. It has been shown that both post-annealing \citep{chikazumi1956} and growth \citep{chikazumi1961} in the magnetic field showing orientation dependency for the single-crystal Py. In an effort to understand such a complication, turned out that Py has negligible magnetostriction i.e.\ its elastic energy is two orders of magnitude smaller than the anisotropic energy \citep{chikazumi1964}. Thus, other effects such as the crystalline anisotropy can easily overcome the magnetostriction. We have already shown that when the tilt deposition competes with a magnetic field, the magnetic axes of polycrystalline films is dictated by the tilt effect \citep{kateb2017}. More recently, we showed that the tilt effect can dictate magnetic axis along the [100] orientations of Py single-crystal \citep{kateb2019epi} while both annealing and growth in the field failed to do so \citep{chikazumi1956,chikazumi1961}. We believe in both single and polycrystalline Py, disordered arrangement of Ni and Fe becomes dominant effect and leads to uniaxial anisotropy in Py. While in an ordered Py, crystalline anisotropy becomes dominant \citep{kateb2019epi}. Since Py exhibits a shrinkage upon ordering, the level of order in a single-crystal can be analyzed by X-ray diffraction (XRD). In polycrystalline films, however, grain size and defect-induced strain effectively contribute to the broadening of XRD peaks and thus limit detection of the atomic order.

In the present study, we aim to understand the origin of uniaxial anisotropy induced by normal deposition with an \emph{in-situ} magnetic field during the growth. In order to minimize the contribution of magnetocrystalline anisotropy, we grow nanocrystalline films. In this case, however, it is not trivial to detect atomic order by XRD as mentioned above. To solve this issue, we utilize resistivity measurement which strongly correlates with the atomic arrangement of Fe and Ni in Py. We show the principal resistivity axes are aligned with magnetic axes of the film which can be only explained by the atomic order.

\section{Experimental Method}
The substrates used here were p-Si (001) with a 100~nm thick layer of thermally grown oxide. After proper cleaning, the substrates were dehydrated at 140~$^\circ$C and their surface modified with HMDS vapor to become hydrophilic before spin-coating of the photoresist. Then the samples were exposed to deep UV and developed. Our depositions were carried out in a UHV ($1.4\times10^{-6}$~mbar base pressure) magnetron sputter system at a pressure of $3.3\times 10^{-3}$~mbar and 150~W. A 4~nm thick Cr layer deposited to increase the adhesion of Py to the substrate and eliminate problems during lift-off. The films were prepared using Py target in normal incidence deposition geometry and utilized a conventional method of applying an \emph{in-situ} magnetic field (70~Oe) at the substrate to induce uniaxial anisotropy in the desired direction. The Hall-bar (0.4$\times$1.6~mm$^2$) and vdP (15$\times$15~mm$^2$) samples were grown simultaneously to make sure there is no difference between them from the magnetic and microstructure viewpoint. Further detail on the preparation can be found in Ref.~\cite{kateb2019}.

The film thickness and grain size were characterized by X-ray reflectivity (XRR) and grazing incidence X-ray diffraction (GIXRD), respectively. To obtain hysteresis loops, we used a high-sensitivity magneto-optical Kerr effect (MOKE) looper. 

\begin{figure}
    \centering
    \includegraphics[width=1\linewidth]{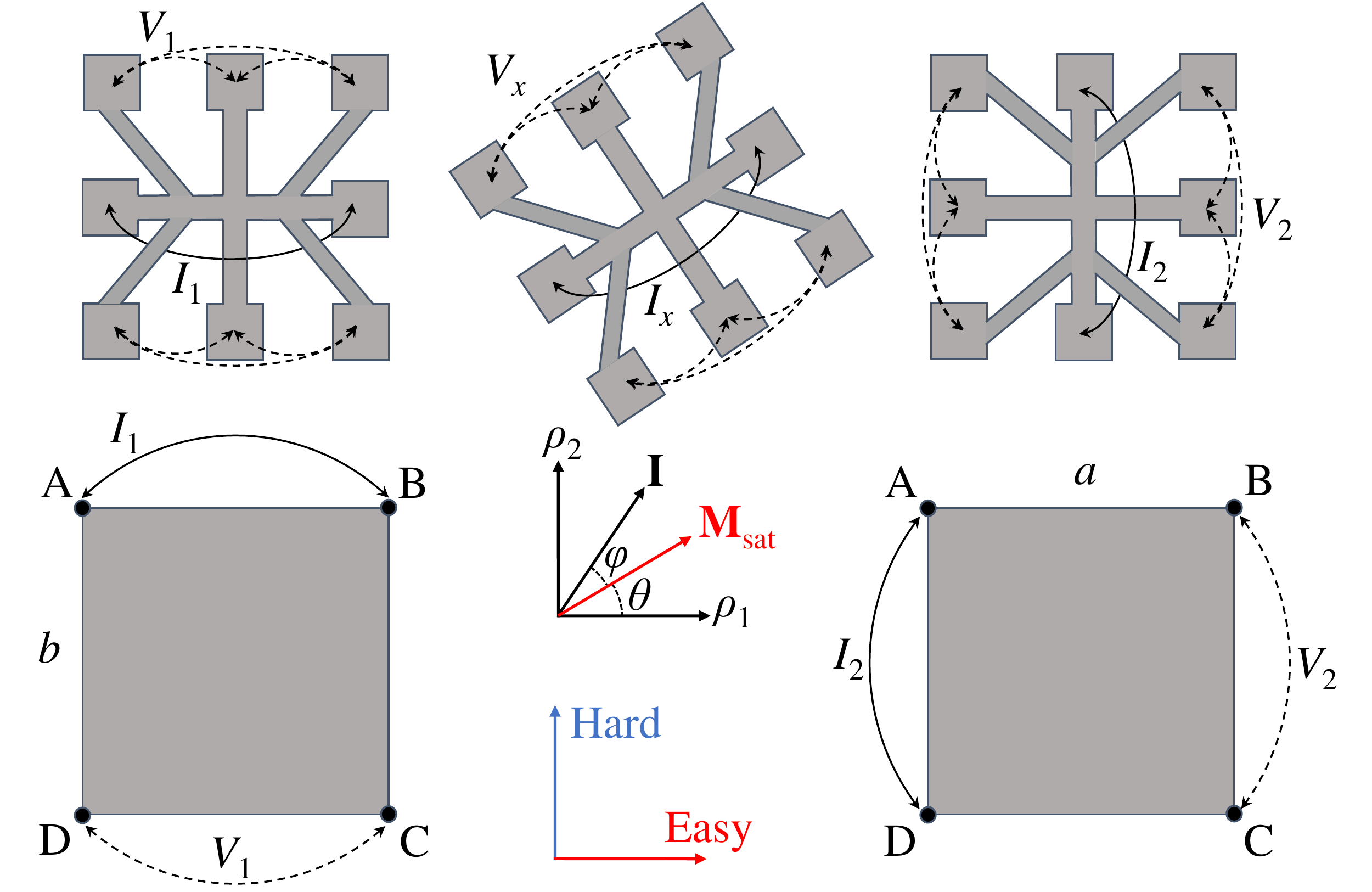}
    \caption{Schematic illustration of resistance measurements along the hard and easy axis for both Hall-bar and vdP methods. The solid and dashed double-sided arrows indicate placements of current and possible arrangements of voltage probes, respectively. $\boldsymbol{\rm I}$ and $\boldsymbol{\rm M}_{\rm sat}$ are the current and saturated magnetization, respectively.}
    \label{fig_scheme2}
\end{figure} 
The resistance was measured by two conventional Hall-bar and vdP methods for comparison which is schematically shown in Fig.~\ref{fig_scheme2}. The vdP method is a simple and flexible technique to probe the resistivity of uniform, continuous thin-films of arbitrary shape \cite{vdPauw1958,Pauw1958}. In the vdP method, four small contacts must be placed on the sample perimeter, not necessarily at the corners labeled A--D in Fig.~\ref{fig_scheme2}. The isotropic resistivity value ($\rho_{\rm{iso}}$) is obtained by the vdP formula:
\begin{equation}
	\exp\left(-\frac{\pi d}{\rho_{\rm{iso}}}R_{\rm AB,CD}\right)+\exp\left(-\frac{\pi
	d}{\rho_{\rm{iso}}}R_{\rm AD,CB}\right)=1
	\label{eq:vdP}
\end{equation}
where $d$ is the film thickness and e.g.\ $R_{\rm AB,CD}$ is the resistance obtained by forcing current through AB and picking up the voltage at the opposite side between CD or \emph{vice versa}.

Since Eq.~(\ref{eq:vdP}) obtained by conformal mapping of a finite sample into an infinite half-plane with contacts along the edge, it should be valid if e.g. an anisotropic rectangle with the lateral dimensions of $a\times b$ is mapped into an isotropic one with $a\times b'$. It has been shown that $\rho_{\rm{iso}}=\sqrt{\rho_1\rho_2}$ is the geometric mean of the principal resistivities \cite{Pauw1958,Price1973}. The individual values of principal resistivities can subsequently be obtained using Eq.~(\ref{eq:rhoratio}):
\begin{equation}
	\rho_{1}=\rho_{\rm{iso}}\sqrt{\frac{\rho_{1}}{\rho_{2}}}
	\label{eq:rhox}
\end{equation}
\begin{equation}
	\rho_{2}=\rho_{\rm{iso}}\sqrt{\frac{\rho_{2}}{\rho_{1}}}
	\label{eq:rhoy}
\end{equation}

Thus, knowing the $\rho_1/\rho_2$ ratio, one can determine principle resistivities. Here, we use Price \cite{Price1973} extension for determining the ratio that requires a rectangular sample with its sides cut along the $\rho_1$ and $\rho_2$. 
\begin{equation}
    \sqrt{\frac{\rho_{1}}{\rho_{2}}}=-\frac{b}{\pi a}\ln\left(\tanh \left[\frac{\pi
    dR_{\rm AD,CB}}{16\rho_{\rm{iso}}}\right]\right)
    \label{eq:rhoratio}
\end{equation}
where $b$ and $a$ are the side lengths of a rectangular sample and $R_{\rm AD,BC}$ is resistance along the $b$ sides as described above. It is worth mentioning that Eq.~\ref{eq:rhoratio} requires knowing the direction of principal resistivities. However, it is possible to determine the whole resistivity tensor in an arbitrary direction \cite{ingvarsson2017} and find the direction of principal resistivities using matrix rotation.

To decouple the effect of magnetization on resistivities, we have rotated $\boldsymbol{\rm M}_{\rm sat}$ to extract $\rho_{\rm{ave}}$. Thus the individual resistivities are expected to behave as below: 
\begin{equation}
    \rho_{\rm long}=\rho_{\bot}+\Delta\rho\cos^2\phi
    \label{eq:rholong}
\end{equation}
here $\phi$ stand for the angle between current and saturated magnetization direction, $\rho_{\rm long}$ is the longitudinal resistivity with respect to the current and $\Delta\rho=\rho_{\|}-\rho_{\bot}$ with $\rho_{\|}$ and $\rho_{\bot}$ being resistivities with $\boldsymbol{\rm M}_{\rm sat}$ parallel and perpendicular to the current direction, respectively. 

The magnetoresistance measurements were done by rotation of an in-plane $\boldsymbol{\rm M}_{\rm sat}$ of $\sim$23~Oe. The strength of the field is enough to saturate the magnetization as it is a few times larger than both coercivity ($H_c$) and anisotropy field ($H_k$). All resistivity measurements were performed at room temperature using a low current density of 3~mA. 
\section{Results and discussion}
The result of XRR measurements (not shown here) fitted according to the Parrat formalism \citep{parratt54:359} to determine film thickness, density and surface roughness. The thickness of the film, which is required to solve Eq.~(\ref{eq:vdP}), is found to be 40~nm. We used the Scherrer equation \citep{langford1978} to estimate the grain size from the (111) GIXRD peak (not shown here) which is found to be $\sim$10~nm. Note that the grain size obtained by the Scherrer equation from GIXRD has been shown to be in agreement with the TEM result for thin Py films \citep{Neerinck1996}. We could not detect any peak shift due to order/disorder \citep{kateb2019epi} since such a small grain size causes considerable peak broadening. Further, we studied texture in the film by polar mapping of (111) plains that indicates the lack of any texture in film. This is the case for the very thin-film in which different crystal planes are still competing and none became dominant and thus grains are essentially equiaxed. Thus, the XRD characterization indicates the film is isotropic.

\begin{figure}
    \centering
    \includegraphics[width=1\linewidth]{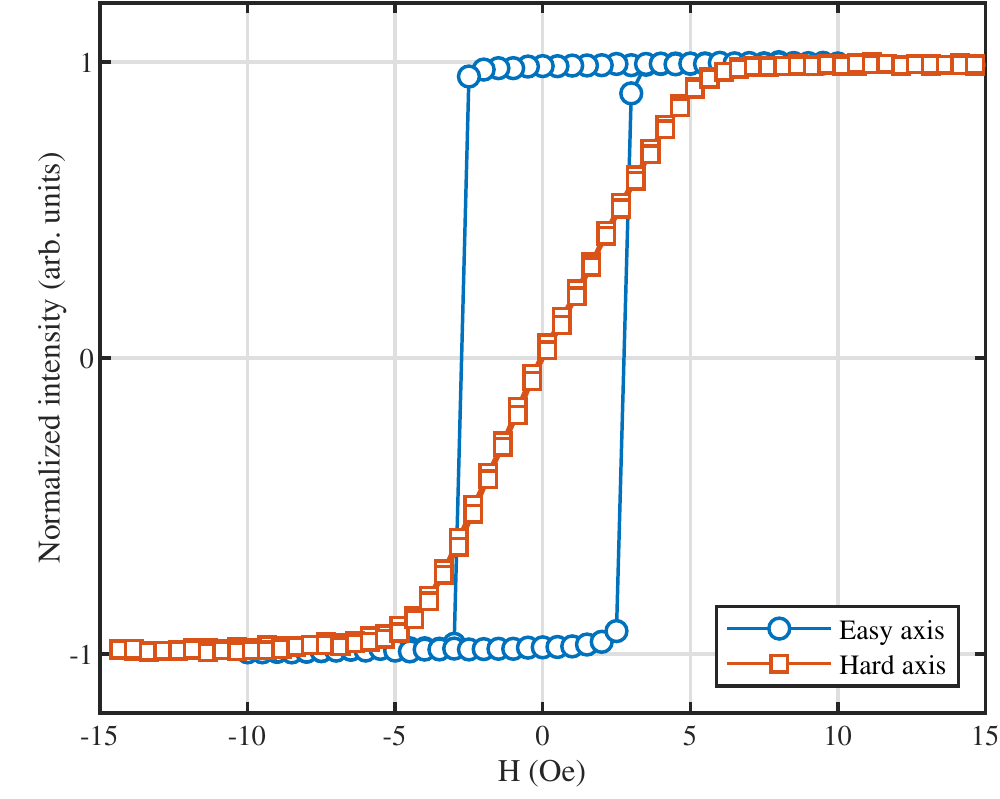}
    \caption{The MOKE response of the as-deposited film along the hard and easy axis.}
    \label{fig:MOKE}
\end{figure}

Fig.~\ref{fig:MOKE} shows the MOKE response of as-deposited films before lift-off. It can be seen that the hard axis presents a completely linear behavior without hysteresis and $H_k$ of 5~Oe. The easy axis presents a square loop with a slight rounding at the corners but still sharp switching and $H_c$ of 2.75~Oe. These indicate the film presents very well-defined in-plane uniaxial anisotropy magnetization induced by applying an \emph{in-situ} magnetic field during the growth. The dimensions of Hall-bars are chosen to be large enough to maintain induced magnetic anisotropic axes after lift-off.

\begin{figure}
    \centering
    \includegraphics[width=1\linewidth]{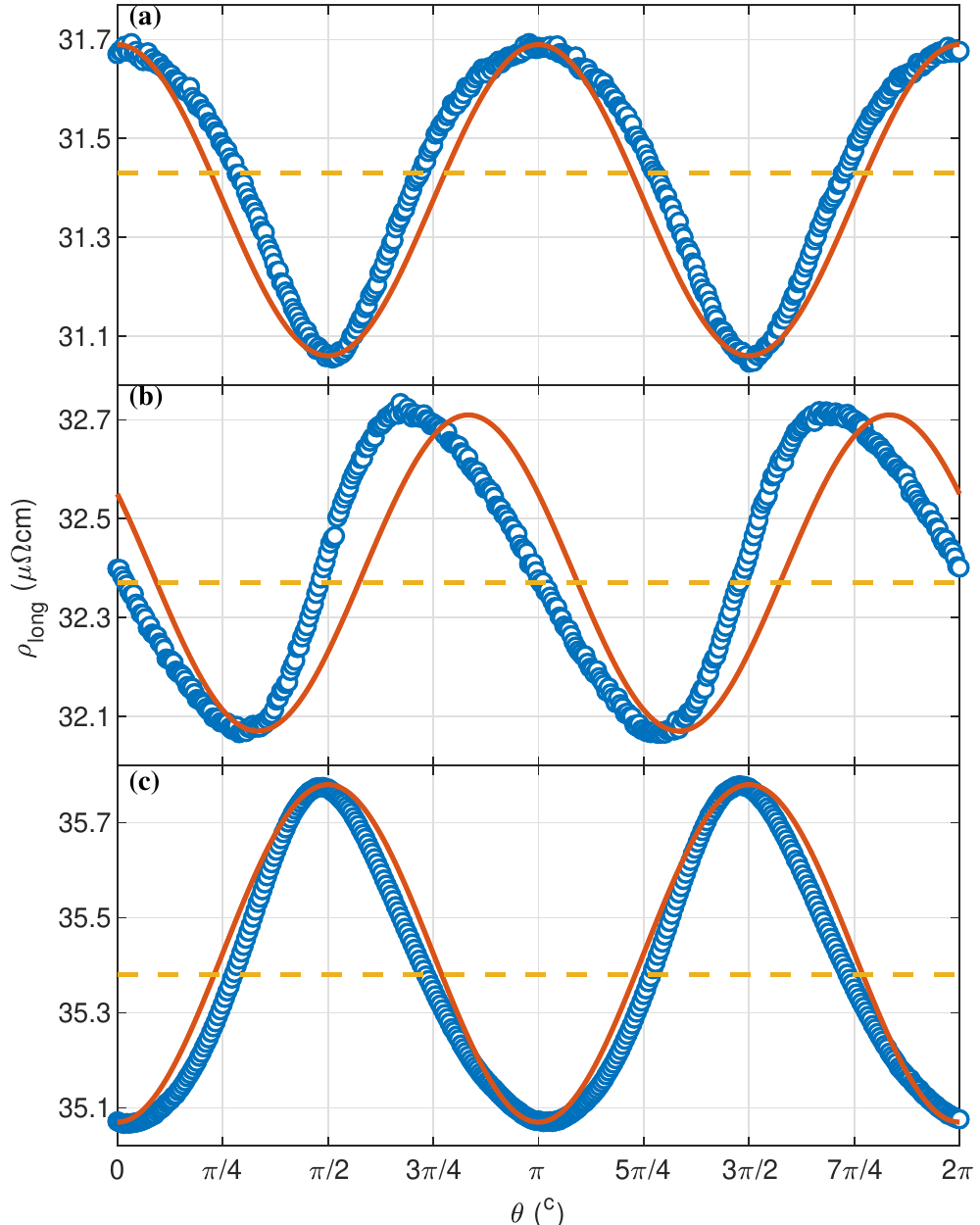}
    \caption{Variation of $\rho_{\rm long}$ with the rotation of $\boldsymbol{\rm M}_{\rm sat}$ for Hall-bars patterned with (a) 0, (b) 30 and (c) 90$^\circ$ with respect to the easy axis. The solid and dashed lines indicate fitting with Eq.~(\ref{eq:rholong}) and the average of $\rho_{\rm long}$, respectively.}
    \label{fig_Hbar}
\end{figure}

Fig.~\ref{fig_Hbar} shows the variation of resistivity with the rotation of magnetization for Hall-bar patterns made with 0, 30 and 90$^\circ$ with respect to the easy axis. Here $\theta$ is defined as the angle between saturated magnetization and easy axis and should not be confused with $\phi$ i.e. the angle between magnetization and current directions (cf.~Fig.~\ref{fig_scheme2}). It can be seen that the absolute value of resistivity and consequently $\rho_{\rm ave}$ increases as we rotate the Hall-bar from 0 to 90$^\circ$. The anisotropic magnetoresistance (AMR) behavior is evident for all cases i.e.\ a maximum resistivity when magnetization is parallel to the current direction and decreases by rotation away from the Hall-bar axis. The AMR measurement by the Bozorth method \cite{Bozorth1946}, is performed by applying $\boldsymbol{\rm M}_{\rm sat}$ parallel and perpendicular to the current direction which is expected to be independent of the direction in a polycrystalline film. Comparing different hall-bars, we observe the difference in both AMR value and the absolute value of resistivity. The result of Eq.~(\ref{eq:rholong}) is also shown for each Hall-bar as a solid line. It can be seen that Eq.~(\ref{eq:rholong}) gives a much better estimation along the easy (0) and hard (90$^\circ$) axes while for the 30$^\circ$ Hall-bar the agreement is quite poor. 

\begin{figure}
    \centering
    \includegraphics[width=1\linewidth]{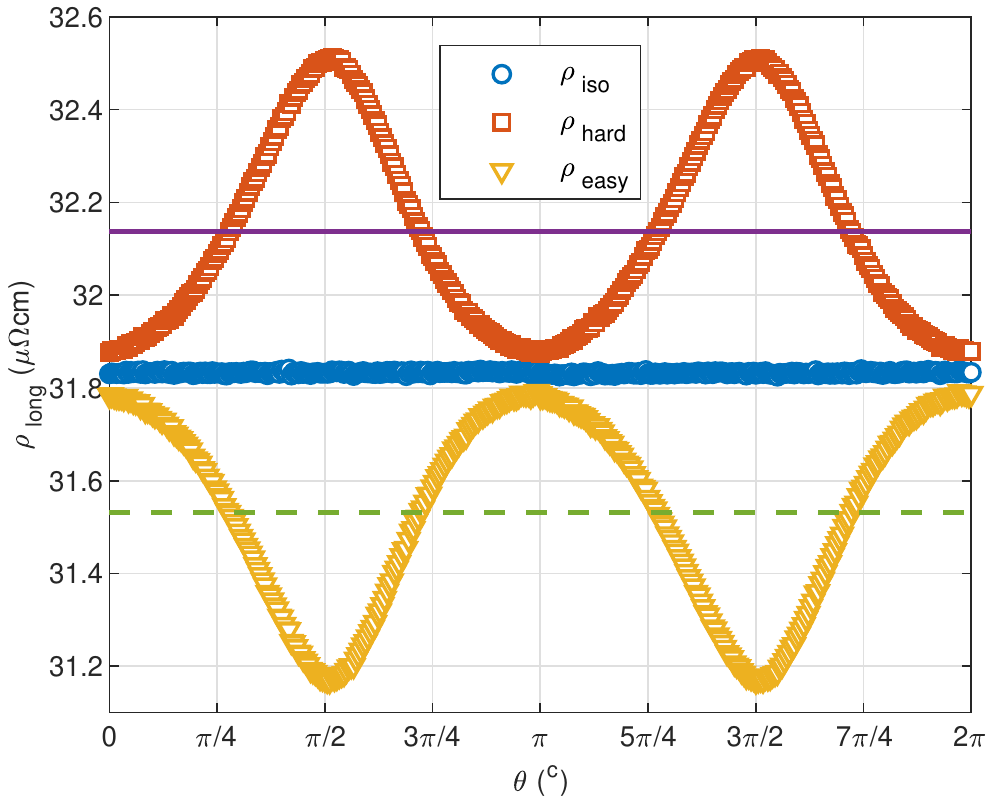}
    \caption{Variation of $\rho_{\rm iso}$ and resistivities along the hard and easy axes due to the rotation of $\boldsymbol{\rm M}_{\rm sat}$ obtained by the extended vdP method. The solid and dashed lines indicate the average of $\rho$ along the hard and easy axis, respectively.}
    \label{fig_5pt}
\end{figure}

Note that the Hall-bars were fabricated large enough to make sure their geometry does not affect initial magnetic anisotropy. Further, we measured anisotropic resistivity using our extended vdP method that can be applied to square-shaped samples \citep{kateb2019}. Fig.~\ref{fig_5pt} shows the variation of different resistivities obtained using the vdP method by rotation of saturated magnetization. It can be seen that the original vdP gives $\rho_{\rm iso}$ which does not change with the rotation of magnetization and thus it is not appropriate for AMR measurements on its own. On the other hand, $\rho_1$ and $\rho_2$ obtained by Eq.~(\ref{eq:rhox}) and (\ref{eq:rhoy}), change symmetrically around the $\rho_{\rm iso}$. This explains why the rotation of saturated magnetization does not change $\rho_{\rm iso}$. It is also clear that both $\rho_1$ and $\rho_2$ are characteristic AMR curves and can be used to determine AMR along the hard and easy axes, respectively. Similar to Hall-bars, different AMR and the absolute value of resistivity are obtained along the hard and easy axis. Again, the $\rho_{\rm ave}$ along the hard axis is higher than that along the easy axis indicating the existence of a correlation between magnetic anisotropy axes and principal resistivities.

In the semiconductor and superconductor community, the resistivity tensor is commonly obtained by patterning and measurement of two Hall-bars along the principal resistivity axes. This is based on the fact that in an anisotropic film, off-diagonal elements of resistivity tensor become zero when measurements are performed along the principal resistivity axes. It is worth mentioning that we measured off-diagonal resistivities to be 5 orders of magnitude smaller and not exactly zero. Once the principal resistivities ($\rho_1$ and $\rho_2$) are determined one can determine resistivity in any direction using matrix rotation. For instance, we can apply this to calculate $\rho_{\rm ave}$ for the 30$^\circ$ Hall-bar from 0 and 90$^\circ$ ones which gives a value of 32.36~$\mu\Omega$cm. This is in close agreement with the $\rho_{\rm ave}$ obtained by resistivity measurement in our 30$^\circ$ Hall-bar. This indicates magnetic anisotropy axes of the sample are aligned with the principal resistivity axes. Note that $\rho_1$ and $\rho_2$ in our scheme are directions with lowest and highest $\rho_{\rm ave}$ obtained by rotation of $\boldsymbol{\rm M}_{\rm sat}$. The latter indicates magnetostriction and internal magnetization cannot explain the alignment of magnetic anisotropy and principal resistivity axes. The only explanation left is that both of these are correlated with the atomic arrangement. The latter includes atomic ordering and pair ordering which can be applied to single-crystal and polycrystalline films, respectively. 

\begin{figure}
    \centering
    \includegraphics[width=1\linewidth]{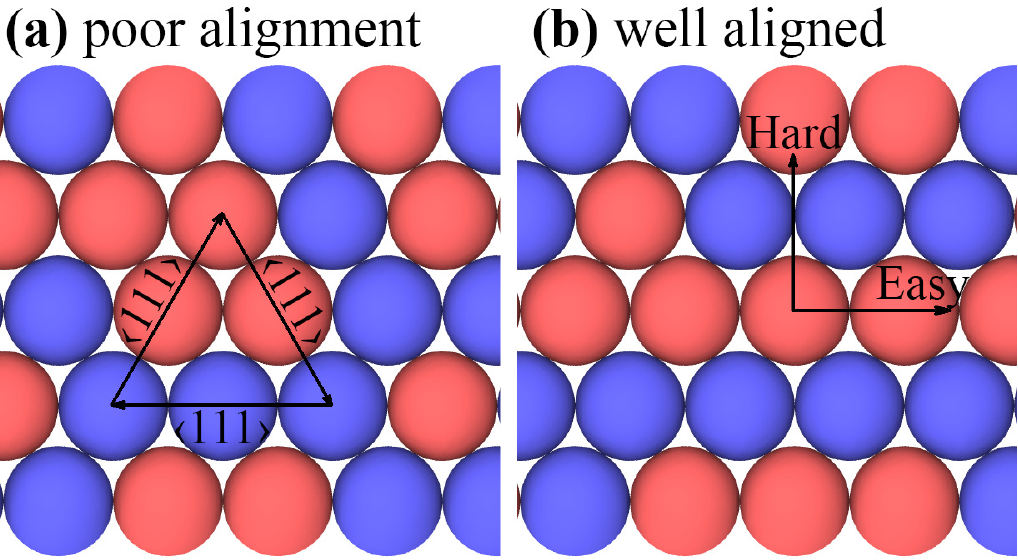}
    \caption{Schematic illustration of atomic order with (a) 3 possible easy orientations within the poor alignment and (b) a dominating easy axis in a well-aligned arrangement.}
    \label{fig_order}
\end{figure}

We have recently shown that an increase in the atomic order towards L1$_2$ Ni$_3$Fe superlattice changes uniaxial anisotropy into four-fold (biaxial) anisotropy and decreases resistivity \citep{kateb2019epi}. In the pair ordering model, the direction of Ni-Ni, Fe-Fe and Ni-Fe pairs determine magnetic anisotropy \citep[p.~339]{cullity1972}. Fig.~5 shows the (111) plane of fcc structure for a binary mixture as an example. Since $\langle111\rangle$ orientations are the easiest directions of fcc Py \citep[p.~199]{cullity1972}, any of the orientations indicated in Fig.~5(a) are a possible easy axis. In the case of poor alignment, the number of each pair is similar in different orientations and none becomes dominating. Thus, the easy axis changes locally and uniaxial anisotropy is an average behaviour over the whole plane \citep{rodrigues2018}. Since the \textit{in-situ} magnetic field during the growth aligns Ni-Ni and Fe-Fe pairs parallel to the field it makes one of the orientations dominating that gives a clear easy axis and uniaxial anisotropy (cf. Fig.~\ref{fig_order}(b)). Thus, along the easy axis electrons are traveling through a more uniform medium where they mainly scatter at the blue/red interface. Such a scatterer, i.e.\ parallel to the transport direction, can be quantitatively described by an approximation of Fuchs \cite{Fuchs1938} model
\begin{equation}
    \rho=\rho_{0}\left(1+\frac{3}{8}\frac{\lambda_{0}}{d}\right)
    \label{eq:Fuchs}
\end{equation}
where $\rho_{0}$ and $\lambda_{0}$ are the bulk resistivity and mean free path of electrons, respectively.
For the transport along the hard axis, one can consider Mayadas and Shatzkes \cite{Mayadas1970} model that includes scatterers perpendicular to the current direction, such as grain boundaries 
\begin{equation}
    \frac{\rho_{0}}{\rho}=3\left[\frac{1}{3}-\frac{\alpha}{2}+\alpha^{2}-\alpha^{3}ln\left(1+\frac{1}{\alpha}\right)\right]
    \label{eq:MS}
\end{equation}
\begin{equation}
    \alpha=\frac{\lambda_{0}}{D}\frac{\Re}{1-\Re}
    \label{eq:alpha}
\end{equation}
where $D$ is the grain size or separation between scatterers and $0<\Re<1$ is the coefficient of grain boundary reflection. Unfortunately, there exists no experimental or theoretical report on the $\Re$ value at the Ni/Fe interface. To give an estimation of $\Re$, in good conductors such as Cu or Ag, a single twinning 
gives $\Re=0.12$--0.17 \citep{feldman2010}. For multiple twinnings, more than 29 and 21\% increase in resistivity is expected for Cu and Ag, respectively. It is also worth mentioning that, although these models were originally developed for thin-films, they are applicable to other objects such as stripes or nanowires, cf.\ \citep{chawla2011}. 

In Fig.~5(b), $d$ in Fuchs \cite{Fuchs1938} model and $D$ in  Mayadas and Shatzkes \cite{Mayadas1970} model are both equal to one atomic layer. Assuming a constant $\lambda_0$, and a reasonable $\Re$ one can conclude that resistivity along the hard axis is higher when similar pairs are aligned with the easy axis.
%


A question that might arise here is why the result of measurements using the Hall-bars and vdP are not the same. Most of critics to vdP focus on the fact that it probes the whole plane of the sample not a specified direction. This has been already answered mathematically for the co-linear and square four-point-probes techniques and we refer the interested reader to e.g. Ref~\cite{miccoli2015}. We considered the source of errors to be patterning in the Hall-bars and size and placement of contacts in the vdP method. Although we have tried to pattern and grow all samples together, there might be little difference between different Hall-bars. But we reduced measurement error in the Hall-bars by utilizing an 8-pad pattern as schematically shown in Fig.~\ref{fig_scheme2}. In such a pattern two pads are utilized for applying
the current (cf.\ the solid double-sided arrow in Fig.~\ref{fig_scheme2}) and 6 remaining pads allow for 6 combinations  of measurements along the Hall-bar (indicated by dashed double-sided arrows). Note that measurements across the Hall-bar include the planar Hall effect. We measured all 6 combinations and rejected those with more than 5\% difference in the averaging. Original vdP requires infinity small contacts for an arbitrary plane \citep{vdPauw1958}. Using four 2~mm contacts, spread into the film, a $\sim$2~\% error was obtained compared to 0.2~mm contacts. The extended vdP is based on rectangle/square geometry that might be affected by contact displacement. For displacing one contact equal to $a/2$ or $b/2$ (cf. Fig.~\ref{fig_scheme2}), i.e. having a contact at sides rather than the corner of the square, an error of 7~\% was produced. Thus, we attribute the difference between Hall-bars and vdP measurement to the patterning.


\section{Conclusion}
Py film without any texture is prepared by sputtering in presence of a magnetic field to induce in-plane uniaxial magnetic anisotropy. It is shown that the $\rho_{\rm ave}$ obtained by rotation of $\boldsymbol{\rm M}_{\rm sat}$ using both Hall-bars and vdP methods is highest along the hard axis and lowest along the easy axis. Thus, it can be concluded that magnetic anisotropy axes and anisotropic resistivity axes are aligned with each other. This is due to the fact that both of these are correlated with the atomic/pair order. In particular, alignment of Ni-Ni and Fe-Fe pairs along the easy axis present a lower scattering rate. While along the hard axis Ni-Fe pairs provide less uniform medium and higher $\rho$. The results indicate it is possible to determine the direction of the magnetic axes by measuring the resistivity tensor. In the latter case, we suggest using extended vdP as a versatile method that can be performed without pattering.

\section*{Acknowledgment}

This work was supported by the Icelandic Research Fund Grant No. 120002023.

\bibliographystyle{elsarticle-num} 
\bibliography{Ref}

\end{document}